\documentclass{article}
\usepackage{amsmath}
\usepackage{graphicx}
\usepackage{hyperref}

\newcommand{\la} {\langle}
\newcommand{\ra} {\rangle}

\title{Testing GPT-4-o1-preview 
on math and science problems: A follow-up study}

\author{Ernest Davis \\ New York University \\ davise@cs.nyu.edu}

\begin{document}
\maketitle

\begin{abstract}
In August 2023, Scott Aaronson and I 
reported the results of testing GPT4 with the Wolfram
Alpha and Code Interpreter plug-ins over a collection of 105 original
high-school level and college-level science and math problems
(Davis and Aaronson, 2023). In September 2024, I tested the recently
released model GPT-4o1-preview on the same collection. Overall I found
that performance had significantly improved, but was still considerably
short of perfect. In particular, problems that involve spatial reasoning
are often stumbling blocks.
\end{abstract}

In August 2023, Scott Aaronson and I
reported the results of testing GPT4 with the Wolfram
Alpha and Code Interpreter plug-ins\footnote{OpenAI stopped supporting 
plug-ins in April 2024. Current versions of GPT4 seem to incorporate the
functionality of Code Interpreter; that is, they can construct Python
programs, execute them, and interpret the results. As far as I know, there is
no longer any way for any supported version of GPT4 to interact with 
Wolfram Alpha.}  over a collection of 105 original
high-school level and college-level science and math problems
(Davis and Aaronson, 2023). 

On September 12, OpenAI (2024) released two preliminary versions, 
``ChatGPT-o1-preview'' and ``ChatGPT-o1-mini'' of a forthcoming product
``ChatGPT-o1''. In their release announcement they claimed, ``o1-preview has 
strong reasoning capabilities and broad world knowledge.'' I tested
ChatGPT-o1-preview (henceforth ''O1'')
over a couple of weeks after its release on the same
collection of problems. This note reports the results of this experiment.

Overall I found
that performance had significantly improved, but was still 
short of perfect. In particular, problems that involve spatial reasoning
are often stumbling blocks.

I will first summarize the results, then discuss the cases
where O1 got the wrong answer or where there is some other 
significant observation to be made. The full collection of problems
with correct answers may be found in the appendices to (Davis and 
Aaronson, 2023). The slightly edited output of GPT-4 with the plugins
from our earlier experment are at an online 
site.\footnote{\url{https://cs.nyu.edu/~davise/papers/GPTPlugInTests/}}

It may be noted that these problems and their correct answers have been
posted since September 2023, so there could have been data contamination.
I doubt that that played a large part in the outcome.

\section{Overall results}

The 105 problems consist of three datasets of problems of somewhat different
flavors, which we called the ``Arbitrary Numerical'' problems, the
``Calculation-Free'' problems, and the ``Motivated Numerical'' problems.
The first two datasets were written by Davis, the third was 
written by Aaronson; they reflect the interests and tastes of their authors.  

Of the 32 ``Arbitrary Numerical'' problems, O1 got 8 
(\#s 7, 10, 15, 17, 18, 19, 30, and 32) wrong and the remaining 24 right.
In our earlier experiment, we gave ChatGPT plus Wolfram Alpha a score of
8.25, with completely correct answers on 6 and .75 partial credit on each
of 3. The remaining 23 were completely wrong.
ChatGPT plus Code Interpreter got 10 completely right and 22 wrong.

Of the 53 ``Calculation-Free'' problems, O1 got five 
(\#s 4-8, and 14)  wrong and the remaining 48 right.
In our earlier experiment, ChatGPT plus Wolfram Alpha got 22 completely wrong, 
3 nearly right, and 28 completely right.
ChatGPT plus Code Interpreter got 18 completely wrong, 1 more wrong than 
right, 3 nearly right, and 31 completely right.

Of the 20 ``Motivated Numerical'' problems O1 got two
(\#s 16 and 17) wrong and the remaining 18 right. 
In our earlier experiment, ChatGPT plus Wolfram Alpha got 5 completely wrong, 
2 partly right, and 13 completely right.
ChatGPT plus Code Interpreter got 5 completely wrong, 3 partly right, and
and 12 completely right.

\section{Discussion of some specific problems}

\subsection{Arbitrary Numerical problems}

{\bf Problem 7} reads: How many total eclipses of the moon were 
there between Jules Verne’s death and Neil Armstrong’s moon landing? 
An exact integer value is required.

O1 correctly determined the dates of Jules Verne's death
and the moon landing, and then enumerated a list of dates between those two
that it claimed came from the information in NASA's ``Lunar Eclipse 
Page''.\footnote{\url{https://eclipse.gsfc.nasa.gov/LEcat5/LEcatalog.html}}
However, the dates that O1 bore no relation to the dates that NASA lists for
the twentieth century.
I was unable to find the source of the list of dates that O1 used.

{\bf Problem 8} reads:  A physical process generates photons 
whose energies follow a random distribution of the following form: 
For positive energy e, the probability density at e is proportional 
to the value of e in a Gaussian distribution with mean 2 Ev and 
standard deviation 0.01 Ev. The probability of a negative value is zero. 
What is the expected value of the wavelength of a photon produced
by this process? (Give the mathematical answer, assuming that the above 
description is exact, and assuming the standard relation between energy 
and wavelength in a photo. The answer is not physically plausible.)

The point of this problem is that, as the wavelength $\lambda$ goes to
0, the probability density $P(\lambda)$ approaches a limit that, though
tiny, is non-zero. Since the frequency is $h/\lambda$, the integral
for the expected value diverges and the expected value is infinite. 
O1 missed that point, and computed the value of the integral omitting the
divergence at 0.

{\bf Problem 12}  reads:
A pendulum is hanging on a 2 meter cord attached to the ceiling 3 
meters above the floor. It is brought to a position 25 degrees from the 
vertical and released. It swings past the bottom and the cord is cut when it is
10 degrees from the vertical on the far side. Then the bob flies through the
air and hits the ground. What is the distance from the point where the bob
is released to the point where it hit the ground?

O1 did this nearly correctly, but took the y-component of the velocity when
the string is done to be negative instead of positive. That is, it failed
to realize that the bob is now swinging {\em upward.}

{\bf Problem 15} reads as follows:
Draw a circle, on the earth's surface, going through Cairo, Peking, and 
Moscow.  Let S be the area of the part of the earth's surface inside the circle 
and let P be the area of the circle in the plane of the circle. What is S/P?.''

O1 simply conjectured that the circle is a great circle, and made the 
calculation on that basis.

{\bf Problem 17} reads as follows:\footnote{In our original tests in 2023,
the problem was mistakenly written 
``Two $^{31}K$ phosphorus nuclei \ldots''. That was corrected
in the test of ChatGPT-4o1-preview.}
Two $^{31}P$ phosphorus nuclei, with no electrons, are isolated in
space, with coordinates $\la$0,0,0$\ra$ and $\la$10,10,10$\ra$ in a coordinate
system whose unit length is 1 Angstrom. What is the instantaneous acceleration
(a vector with unit length of Angstrom/sec$^{2}$)
of the nucleus at the origin due to the electrostatic force?

The answer that O1 output corresponded to the nucleus at the origin
accelerating {\em toward\/} the other nucleus; that is, it assumed an
attractive rather than a repulsive force. The magnitude of the force was also
somewhat incorrect due to a miscalculation.

{\bf Problem 18} reads as follows:
An irregular (house-shaped) pentagon has vertices numbered 1 through 5 in
order. The pentagon has right angles at vertices 1, 3, and 4, and 135-degree
angles at 2 and 5. Side 2-3 and 4-5 have length 1 and side 3-4 has length 2.
The pentagon is placed on a planar coordinate system so that the numbering of
the vertices is in clockwise order, vertex 3 is at the origin, and vertex 5
is on the positive y-axis. What are the coordinates of vertex 1?

O1 entirely misinterpreted the problem. The geometric calculation it carried
out referred to numbers in the problem as asked, but the geometry was entirely
incorrect.

{\bf Problem 19} reads as follows:
Consider a cube with unit length sides, where the vertices of one
face are numbered A..D in counterclockwise order, as viewed from the center
of the cube;  the vertices of the opposite face are named E to H; and there are
edges AE, BF, CG, and DH.
Rotate the cube so that vertex A is at the origin,
vertex G is on the positive z axis, and vertex B is in the x-z plane, with
positive x coordinate.
What are the coordinates of vertex E?

O1 ignored the constraint that vertex B is in the x-z plane.

{\bf Problem 20} reads as follows:
Joe and Jim each have a bank account which they started on January 1,
2000. Joe started his account with with \$1000; Jim started his with \$900.
Joe's account pays 5\% every December 31, and he
adds an additional \$500 every January 1. Jim's account pays 10\% every
December 31. When will Jim's account have more money in it than Joe's?

O1 made an off-by-one error.

{\bf Problem 30} reads as follows:
A satellite in a circular geosynchronous orbit passes directly above the
North and South Poles. When it crosses the North Pole, its velocity is in the
plane of the $0^{\circ}$ circle of longitude. At what longitude 
does it pass directly above a point in the Tropic of Cancer?
(An answer will be marked correct if it is within 3 degrees of the correct
answer.)
Assume that the satellite is moving in a closed orbit around the Earth
and that the only influence on the satellite's motion  is the
Earth's gravity. Assume that the Earth is a perfect sphere.
Ignore the revolution of the Earth around the sun,
but do not ignore the rotation of the Earth around its axis.

O1 ignored the rotation of the earth in answering this question.

{\bf Problem 32} reads as follows:
A satellite orbits the earth in a circular orbit. It passes directly over the
North and South poles and completes an orbit every 14 hours 40 minutes.
On one orbit going southward it was directly above the earth location
$40^{\circ}$ N,
$10^{\circ}$ W at 1:00 PM EST. At what time will it next cross the
plane that contains
the circle of latitude $40^{\circ}$ N, and what will be its longitude?
(An answer will be marked correct if the time is within 5 minutes of the
correct answer, and the longitude is within 3 degrees.)
Assume that the satellite is moving in a closed orbit around the Earth
and that the only influence on the satellite's motion  is the
Earth's gravity. Assume that the Earth is a perfect sphere.
Ignore the revolution of the Earth around the sun,
but do not ignore the rotation of the Earth around its axis.

O1 failed to understand the difference between ``being directly above an earth
location at $40^{\circ} N$'' vs. ``being in the plane that contains the circle 
of longitude $40^{\circ} N$.''

\subsection{Calculation-Free Problems}
{\bf Problems 1-8} have the following form:
\begin{quote}
An astronaut is standing [in the Sea of Tranquility/on the far side of the
moon] during what on earth is called a total [lunar/solar]
eclipse. They are looking in the direction of the [earth/sun]. 
What they see is: \\
A. The surface of the moon, illuminated by earth light. \\
B. The night side of the earth, occluding the sun. \\
C. The surface of the moon, illuminated only by starlight. \\
D. The surface of the moon, illuminated by the sun. \\
E. The sun. \\
F. The day side of the earth, with a small circular shadow moving quickly over it. \\
G. The night side of the earth. The sun is somewhere else entirely. \\
H. A starry sky. Neither the sun, the earth, or the surface of the moon is in the field of view. \\
\end{quote}

The eight problems correspond to the eight possible combinations of the above
options. O1 got the correct answers for the problems where the astronaut is
in the Sea of Tranquility, but gave the same answer for the corresponding
problems where the astronaut is on the dark side of the moon. In effect, it
failed to realize that the moon would occlude the astronaut's view of earth; and,
in the case of an astronaut looking at the sun during a solar eclipse, that
the moon would no longer occlude their view of the sun. 

{\bf Problem 14} reads as follows: 
Joe says that he lives 1000 miles from Walden Pond, that Beth lives 
1000 miles from Lake Michigan, and that he and Beth live 10 miles apart.
Is it possible that Joe is telling the truth? Answer "Yes" or "No".

O1 answered, incorrectly, "No". It gave no explanation, following its 
instructions.
Problems 9 through 15 are problems of a similar flavor, which effectively
require the answerer to determine how the triangle inequality applies in situation
with extended objects. Certainly a success rate of 6 out of 7 does not 
statistically exclude
the null hypothesis of random guessing (the probability is 1/16).

{\bf Problems 21-32} present a triple of cities, such as Caracas, Venezuela; 
Amarillo, Texas; and Quebec, Quebec; and asks whether a cycle in that
order is clockwise or counterclockwise. O1 answered all these correctly by
looking up the latitude and longitude of the cities involved, and applied
the ``Shoelace'' formula for signed areas directly to the latitudes and 
longitudes, as if they were drawn on a flat map split at the $180^{\circ}$
longitude line.
(To be
precise, I only tested three of the examples; since O1 had found this 
method of answering the questions, it seemed clear that barring some very strange
error, it would continue answering all of them correctly.) It would be interesting
to know {\em how\/} O1 found this somewhat obscure procedure --- is it somewhere
laid out explicitly in the training set, or did O1 combine other procedures
to arrive at it --- but that would probably be very hard to determine.

There are other kinds of questions that would probe the depth of an AI's 
comprehension of clockwise vs. counterclockwise rotation:
\begin{itemize}
\item Use a cycle where O1's method will give the wrong answer. E.g. 
``Does the cycle from San Francisco to Tokyo to Perth back to San Francisco
go clockwise or counterclockwise?''

\item Tracing along the borders of countries or states: 
``A traveller starts out at the
place where France and Belgium meet the English Channel. They then travel along 
the Channel and Atlantic coast of France, then along the border with Spain, then
along the Mediterranean, then along the eastern border of France until they
return to their starting point. Is this cycle clockwise, counterclockwise,
or self-intersecting?''

\item Using well known shapes. ``Someone is drawing the digit 6 on paper. They
start at the loose tail on the top, then travel smoothly around the shape until
the entire digit is drawn. Is the cycle at the end drawn in a clockwise or 
counterclockwise direction?''
\end{itemize}

However, since these were not part of our original dataset, I did not run
these in this experiment.

\subsection{Motivated Numerical Problems}

{\bf Problem 16} reads: 
Compute the Shannon entropy of a positive integer n that's chosen
with probability Pr[n] = $6/(\pi^2 \cdot n^2)$ to 3 digit accuracy. 
Please include all calculations.

In doing the multistep calculation, O1 at one point dropped a factor of 
$\pi^{2}$, and thus miscalculated the answer.

{\bf Problem 17} reads: 
Assume that IQs are normally distributed, with a mean of 100 and a
standard deviation of 15.  For which n does the number of people with
IQ n exceed the number of people with IQ n+1 by the maximum amount?
Please include all calculations.

O1's answer was the reverse: the point where the number of people with IQ
n+1 exceeds the number of people with IQ n by the maximal amount. 
  
\section{Conclusion}
On these collection of problems, GPT-4-o1-previews performance was very much
stronger than the performance in August 2023 of GPT4 with either the
Wolfram Alpha or the Code Interpreter plug-ins. It is probably comparable,
on these datasets,
to a strong math or physics major who has access to Wikipedia to look up
the physical constants and geographical information and to a calculator to
do the computations. It is still not perfect on these datasets. In particular,
spatial reasoning is a point of weakness; most of the problems that it got
wrong were either purely geometric or involved a failure of spatial reasoning
in solving a physics problem. 

\subsection*{Acknowledgements} Thanks to Scott Aaronson for feedback.

\section*{References}
Ernest Davis and Scott Aaronson (2023). ``Testing GPT-4 with Wolfram Alpha 
and Code Interpreter plug-ins on math and science problems.''  arXiv 2308.05713.
\url{http://arxiv.org/abs/2308.05713}.

OpenAI (2024). Blog. \url{https://openai.com/o1/}

\end{document}